\documentclass[aps,prl,twocolumn,superscriptaddress,longbibliography]{revtex4-1} 
\pdfoutput=1
\usepackage[version=3]{mhchem} 
\usepackage[dvips]{epsfig}
\usepackage{graphicx}  
\usepackage{dcolumn}   
\usepackage{bm}        
\usepackage{amssymb}   

\begin{document}

\author{Marcelo Wu}
\affiliation{Institute for Quantum Science and Technology, University of Calgary, Calgary, Alberta, T2N 1N4, Canada}
\affiliation{National Institute for Nanotechnology, Edmonton, Alberta, T6G 2M9, Canada}

\author{Aaron C. Hryciw}
\affiliation{National Institute for Nanotechnology, Edmonton, Alberta, T6G 2M9, Canada}

\author{Chris Healey}
\affiliation{Institute for Quantum Science and Technology, University of Calgary, Calgary, Alberta, T2N 1N4, Canada}
\affiliation{National Institute for Nanotechnology, Edmonton, Alberta, T6G 2M9, Canada}

\author{David P. Lake}
\author{Harishankar Jayakumar}
\affiliation{Institute for Quantum Science and Technology, University of Calgary, Calgary, Alberta, T2N 1N4, Canada}

\author{Mark R. Freeman}
\affiliation{National Institute for Nanotechnology, Edmonton, Alberta, T6G 2M9, Canada}
\affiliation{Department of Physics, University of Alberta, Edmonton, Alberta, T6G 2E9, Canada}
\author{John P. Davis}
\affiliation{Department of Physics, University of Alberta, Edmonton, Alberta, T6G 2E9, Canada}

\author{Paul E. Barclay}
\email{pbarclay@ucalgary.ca}
\affiliation{Institute for Quantum Science and Technology, University of Calgary, Calgary, Alberta, T2N 1N4, Canada}
\affiliation{National Institute for Nanotechnology, Edmonton, Alberta, T6G 2M9, Canada}

\title{Dissipative and Dispersive Optomechanics in a Nanocavity Torque Sensor}

\begin{abstract}
Dissipative and dispersive optomechanical couplings are experimentally observed in a photonic crystal split-beam nanocavity optimized for detecting nanoscale sources of torque. Dissipative coupling of up to approximately $500$ MHz/nm and dispersive coupling of $2$ GHz/nm enable measurements of sub-pg torsional and cantilever-like mechanical resonances with a thermally-limited torque detection sensitivity of 1.2$\times 10^{-20} \text{N} \, \text{m}/\sqrt{\text{Hz}}$ in ambient conditions and 1.3$\times 10^{-21} \text{N} \, \text{m}/\sqrt{\text{Hz}}$ in low vacuum. Interference between optomechanical coupling mechanisms is observed to enhance detection sensitivity and generate a mechanical-mode-dependent optomechanical wavelength response.
\end{abstract}

\maketitle


Optical measurement and control of mechanical vibrations are at the heart of many technological and fundamental advances in physics and engineering, from sensitive displacement and force detection \cite{ref:rugar2004ssd, ref:arcizet2006hso, ref:li2009bat, ref:anetsberger2010mnm, ref:gavartin2012aho, ref:sun2012fdc, ref:krause2012ahm, ref:liu2012wcs, ref:kim2013nto}, to proposed observation of gravitational waves \cite{ref:ligo2011gwo} and studies of the quantum properties \cite{ref:meekhof1996gnm} of massive objects \cite{ref:marshall2003tqs, ref:chan2011lcn, ref:teufel2011scm}.  Nanophotonic implementations of cavity optomechanical systems \cite{ref:aspelmeyer2013co} localize light to subwavelength volumes, enhancing optomechanical coupling between photons and phonons of nanomechanical structures \cite{ref:eichenfield2009oc, ref:gavartin2011oct, ref:sun2012fdc}.  Harnessing this optomechanical interaction has enabled milestone experiments, including ground-state cooling \cite{ref:chan2011lcn, ref:safavinaeini2012oqm}, mechanical squeezing of light \cite{ref:safavinaeini2013sls}, and optomechanically induced transparency \cite{ref:weis2010oit, ref:safavi-naeini2011oit}.  Typically, optomechanical coupling arises in cavity optomechanical systems from a dispersive dependence of the nanocavity resonance frequency on the nanocavity geometry, which is modulated by mechanical excitations.  In this paper, we demonstrate that dissipative optomechanical coupling, where mechanical excitations modulate the nanocavity photon lifetime, can also play a crucial role in the optical transduction of nanomechanical motion. In particular, we demonstrate the dissipative-enhanced optomechanical readout of a cantilever integrated directly within a nanocavity and realize an optomechanical torque detector whose sensitivity of $\sim 1.3 \times 10^{-21} \text{N}\, \text{m}/\sqrt{\text{Hz}}$ promises to significantly advance detection of phenomena in studies of nanomagnetic \cite{ref:davis2010nrt, ref:burgess2013qmd} and mesoscopic \cite{ref:bleszynskijayich2009pcn} condensed matter systems, optical angular momentum \cite{ref:yao2011oam}, and magnetometry \cite{ref:forstner2012com}.  We also observe interference between dissipative and dispersive coupling mechanisms, which reveals details of the nature of the nanomechanical motion and may open new avenues in optomechanical control \cite{ref:elste2009qni, ref:huang2010rcb, ref:xuereb2011dom, ref:weiss2013qll, ref:tarabrin2013adb}.

\begin{figure}[htb]
\begin{center}
\epsfig{figure=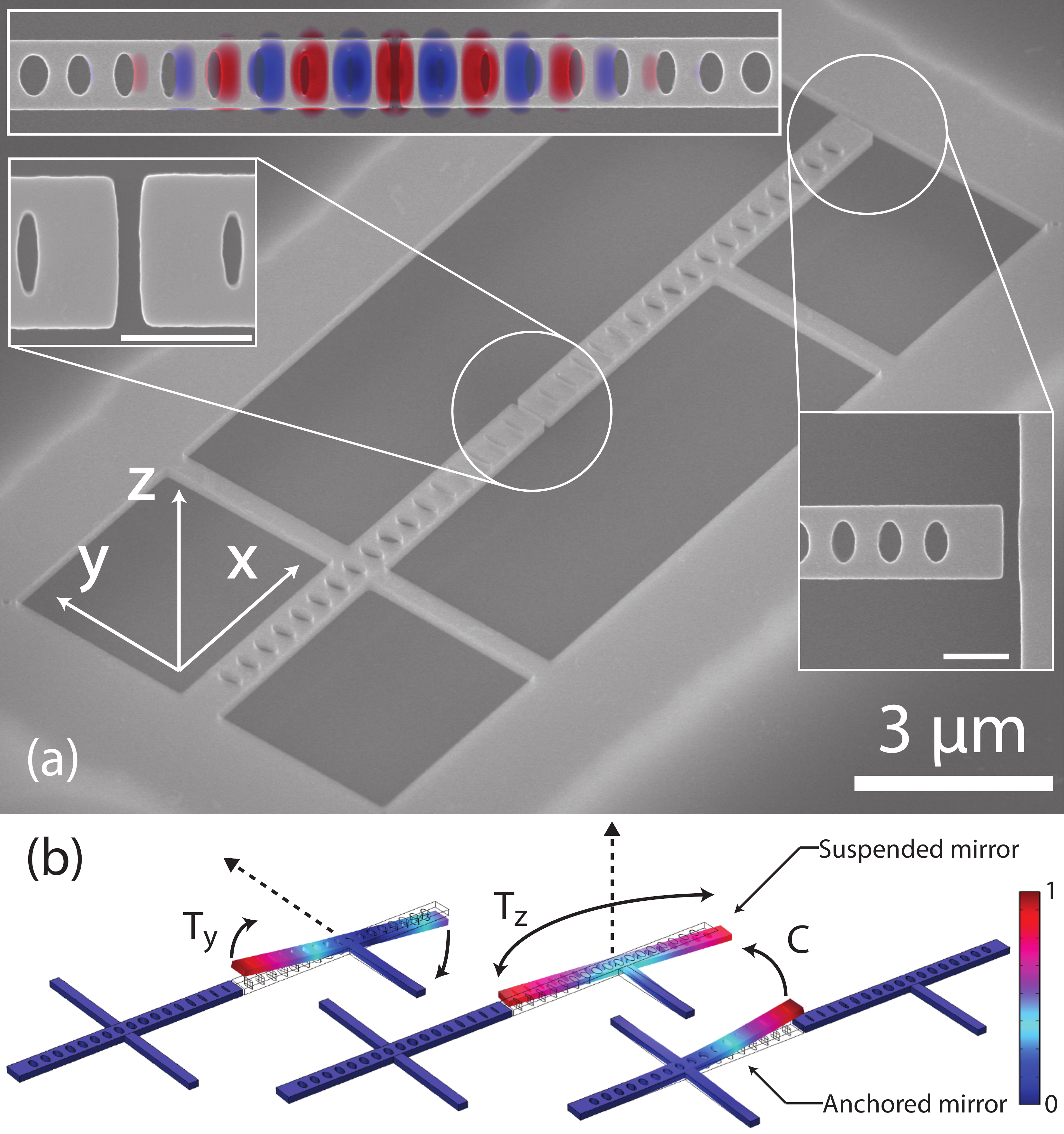, width=1\linewidth}
 \caption{(a) Scanning electron micrograph of a split-beam nanocavity. Top left inset: Top view of the nanocavity overlaid with the field distribution ($E_y$) of the optical mode. Left inset: 60-nm-wide nanocavity central gap. Right inset: Gap separating the suspended nanobeam from the device layer. Inset scale bars: 500 nm. (b) Displacement fields of split-beam nanocavity mechanical modes of interest. Dotted arrows indicate the position and direction of torque for efficient actuation.}
\label{fig:Optics}
\end{center}
\end{figure}


The nanocavity optomechanical system studied here, an example of which is shown in  Fig.\ \ref{fig:Optics}(a), provides a unique platform for studying dispersive and dissipative optomechanical couplings and their impact on sensing and measurement. These photonic crystal ``split-beam'' nanocavities support high optical quality factor ($Q_o$) modes localized between two cantilever nanomechanical resonators, which are patterned to also serve as optical ``mirrors."   The mirrors can move independently and support mechanical resonances whose properties can be customized through design of their mechanical supporting structure. In the device under study, one of the mirrors is suspended by a single mechanical support. Mechanical resonances of this mirror can be efficiently actuated by coupled  sources of torque, as illustrated in Fig.\ \ref{fig:Optics}(b), potentially allowing sensitive readout of a variety of nanomagnetic and mesoscopic  systems \cite{ref:davis2010nrt, ref:burgess2013qmd, ref:bleszynskijayich2009pcn, ref:forstner2012com, ref:yao2011oam}.

\begin{figure}[tb]
\begin{center}
\epsfig{figure=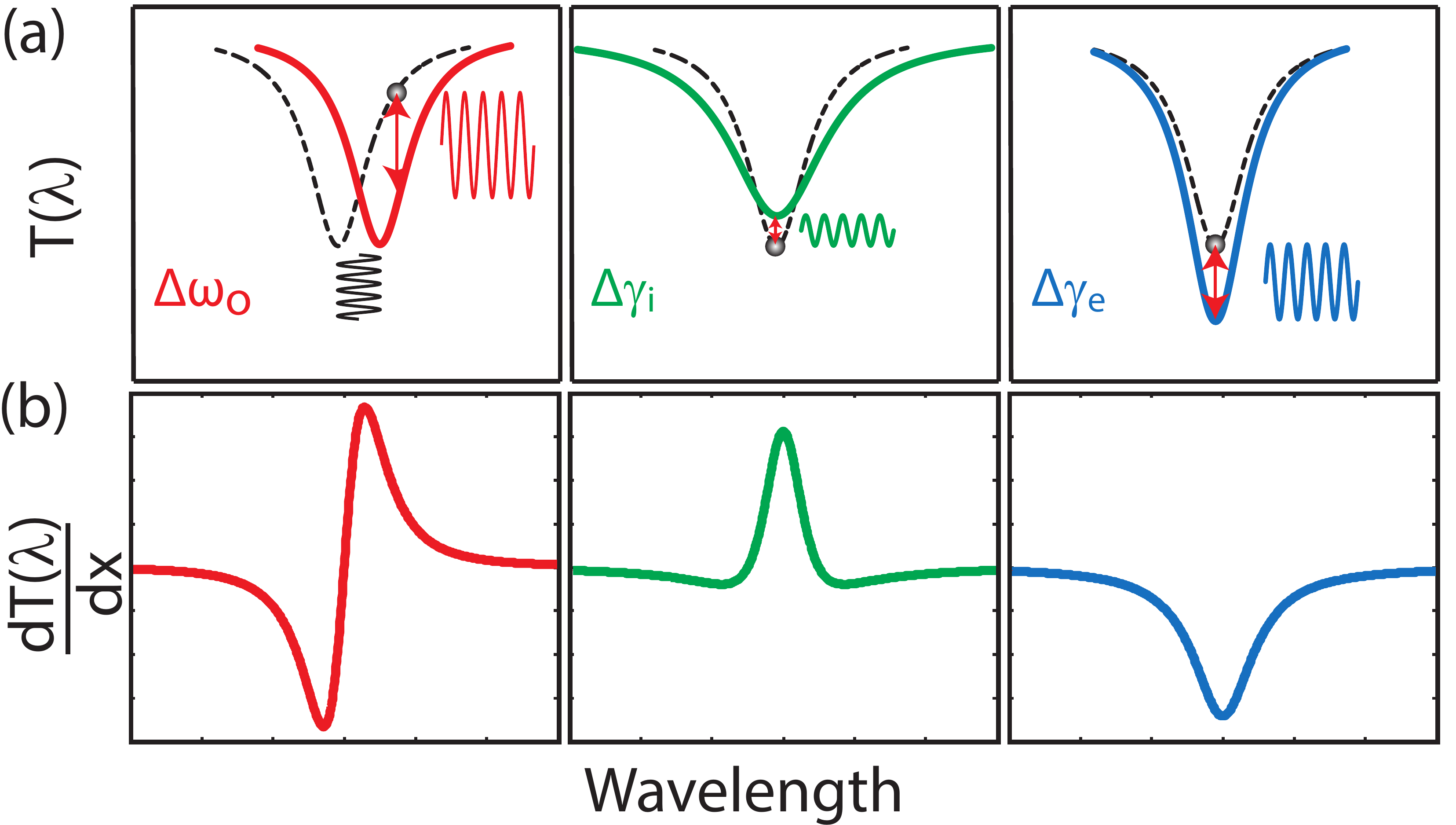, width=1\linewidth}
 \caption{Illustration of the effect of mechanical displacement on the optical response of a cavity with (left) dispersive, (center) dissipative intrinsic, and (right) dissipative external optomechanical coupling. (a) Change in the resonance line shape. (b) Amplitude of the optomechanically actuated signal.}
\label{fig:OptomechComp}
\end{center}
\end{figure}


Optomechanical coupling in split-beam nanocavities is strongest when mechanical motion of the mirrors modifies the nanocavity mirror gap, effectively changing the nanocavity length, resulting in a dispersive coupling to the nanocavity optical frequency $\omega_o$. A more striking property of split-beam nanocavities is the strong dependence of the nanocavity internal photon decay rate $\gamma_i$ on mirror gap, resulting in dissipative optomechanical coupling \cite{ref:elste2009qni, ref:huang2010rcb, ref:xuereb2011dom, ref:weiss2013qll, ref:tarabrin2013adb}. Additional dissipative coupling arises when the motion of the  mirror  modulates the nanocavity external photon decay rate $\gamma_e$ into an external coupling waveguide.  These interactions can be probed by monitoring fluctuations in the transmission $T(\lambda)$ of a waveguide coupling light into and out of the nanocavity, as illustrated in Fig.\ \ref{fig:OptomechComp}(a). For cavity-optomechanical systems operating in the sideband-unresolved regime, a shift $dx$ in the mechanical resonator position modifies $T$ by,
\begin{equation}
dT = \left ( g_{\text{OM}}\frac{\partial T}{\partial \omega_o} + g_i \frac{\partial T}{\partial \gamma_i} + g_e \frac{\partial T}{\partial \gamma_e} \right ) dx,
\label{eq:transmission}
\end{equation}
where $g_\text{OM} = d\omega_o/dx$ is the dispersive optomechanical coupling coefficient, and $g_i = d\gamma_i/dx$ and $g_e = d\gamma_e/dx$ are the intrinsic and external dissipative coupling coefficients, respectively.  The derivatives in Eq.\ \eqref{eq:transmission} can be derived from cavity--waveguide coupled mode theory, and are given in the Appendix. A key feature is that $|{\partial T}/{\partial \gamma_{i,e}}|$ are unipolar, whereas $|{\partial T}/{\partial \omega_o}|$ is bipolar, as illustrated in Fig.\ \ref{fig:OptomechComp}(b).  Interference between these terms can result in an asymmetric optomechanical wavelength response $dT(\lambda)/dx$ with respect to detuning $\Delta\lambda = \lambda - \lambda_o$.  Maximum contributions to $dT$ from dispersive, dissipative intrinsic, and dissipative external optomechanical coupling mechanisms scale with $Q_o/\omega_o\{(1-T_o)g_\text{OM}, \, 4(1-T_o) g_i,\, 8 T_o g_e\}$, and occur when $\Delta\lambda = \{\delta\lambda/2,0,0\}$, respectively, where $T_o = T( \lambda_o)$ and $\delta\lambda = \lambda_o/Q_o$.  Notably, transduction via $g_e$ does not vanish when $T_o \to 1$, i.e., for undercoupled nanocavities ($\gamma_e \ll \gamma_i$).  However, $g_e$ itself does vanish as the fiber taper moves further away from the cavity resulting in degradation of the transduction signal.


The critical role of optomechanical coupling for a wide class of sensing applications is revealed by the minimum force detectable by a cavity optomechanical system \cite{ref:forstner2012com}:
\begin{equation}
F_\text{min}(\omega)= \bigg [ \frac{4k_B T_e m \omega_m}{Q_m} + \frac{S_n}{[G(\lambda) |\chi(\omega)| ]^2 } \bigg ] ^{1/2}.
\label{eq:MinSensitivity}
\end{equation}
$F_\text{min}(\omega)$ describes the minimum actuating force required to obtain a unity signal-to-noise ratio in the presence of thermal noise and technical measurement noise. The lower bound on $F_\text{min}(\omega)$ is fixed by thermal fluctuations of the mechanical resonator, given by the first term in Eq.\ \eqref{eq:MinSensitivity}, and determined by the mechanical resonator frequency $\omega_m$, effective mass $m$,  mechanical quality factor $Q_m$, and temperature $T_e$, where $k_B$ is Boltzmann's constant.  Overcoming technical noise $S_n$ requires a combination of large optomechanical gain $G(\lambda)$ and driving the system at frequency $\omega$ where the  mechanical susceptibility, $\chi(\omega) = [m (\omega^2-\omega_m^2-i\omega \omega_m / Q_m)]^{-1}$, is large.  In many cavity optomechanical systems, including the split-beam nanocavities studied here,  $F_\text{min}(\omega)$ is thermally limited at room temperature, where the thermal phonon population exceeds $10^6$ for MHz-frequency mechanical resonators. In such systems, further reducing the effects of technical noise is advantageous to allow sensitive off-resonance detection, to improve the measurement resolution, and to enhance the ultimate device sensitivity in the case of low-$T_e$ operation. In the measurements presented below, dominant contributions to $S_n$ are from laser noise and photon shot noise, followed by detector noise. Note that backaction noise is not included in Eq.\ \eqref{eq:MinSensitivity},  due to its negligible effect in the regime of operation studied here.

The optomechanical gain, $G(\lambda) =\eta g_ti |dT/dx(\lambda)| P_i$, is determined by Eq.\ \eqref{eq:transmission}, and for a given waveguide input power $P_i$  waveguide transmission efficiency $\eta$, and photodetector gain $g_{ti}$, can be increased through large $g_{\text{OM},i,e}$, high $Q_o$, and optimally tuning $\lambda$ within the nanocavity optical mode linewidth.  As discussed above, dissipative external coupling can play an important role in maximizing $G$. However, dissipative coupling is often small compared to dispersive contributions and, to date, has only been reported experimentally in hybrid cavity-nanomechanical systems where $g_e \sim 10 -- 20$ MHz/nm \cite{ref:li2009rco, ref:anetsberger2011coc}.  In the split-beam nanocavities studied below, measurements indicate that $g_\text{OM} \sim$ 2 GHz/nm,  $g_i \sim 300 - 500$ MHz/nm and $g_e \sim 2 - 3$ MHz/nm.

\begin{figure}[!ht]
\epsfig{figure=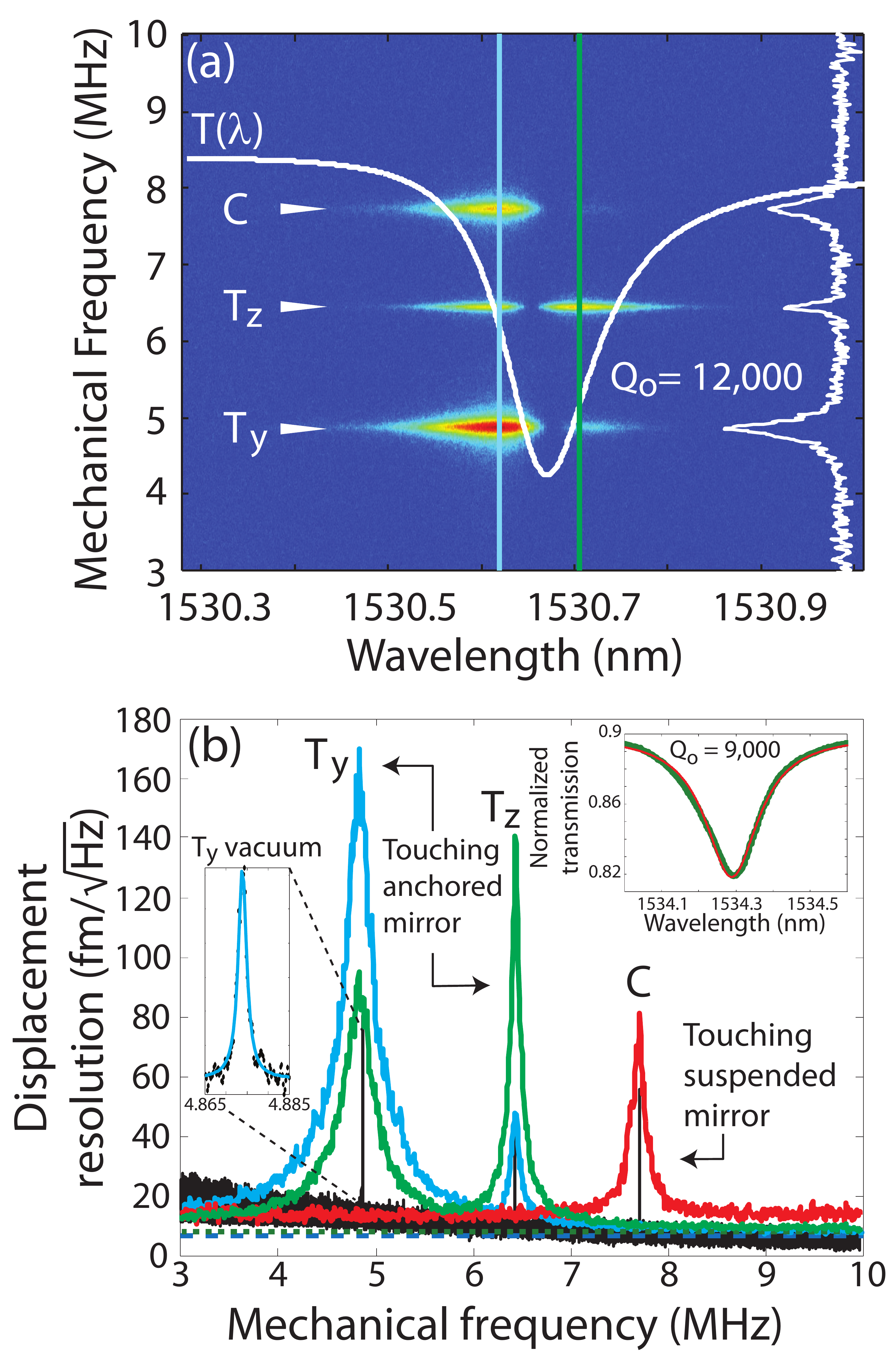, width=1\linewidth}
 \caption{ (a)  $\bar{S}_\text{VV}(\lambda,\omega)$ in ambient conditions, with the fiber taper hovering approximately $300$ nm above the nanocavity; $T(\lambda)$ is superimposed in white. Right side in white: $\bar{S}_\text{VV}(\lambda_b,\omega)$ at $\lambda_b$ indicated by the blue line. (b) Blue (green) data: Calibrated displacement spectrum, $S_{xx}^{1/2}$, of T$_y$ (T$_z$), when the fiber taper is touching the anchored mirror, with $\lambda$ set at the blue (green) line in (a). Dotted lines indicate noise floor. Red data: Uncalibrated displacement spectrum of $C$ with the taper touching the suspended mirror. Black data: Vacuum measurement of the displacement spectrum, uncalibrated. Left inset: Highlight of T$_y$ ($Q_m = 1800$)  in vacuum. Top right inset: $T(\lambda)$ with fiber-touching device.}
\label{fig:OptoMech}
\end{figure}


The split-beam nanocavity devices studied here are fabricated from silicon-on-insulator chips consisting of a 220-nm-thick silicon (Si) layer on a 3-$\mu$m-thick silicon-dioxide (SiO$_2$) layer. Using electron-beam lithography, reactive-ion etching, and a hydrofluoric acid undercut, pairs of cantilever photonic crystal mirrors are defined, with a 60-nm gap between them and another at one mirror end. The resulting split-beam photonic crystal nanocavities, whose design is described in Ref.\ \cite{ref:hryciw2013ods}, support high-$Q_o$ optical modes with $\omega_o/2\pi \sim 200$ THz ($\lambda_o \sim 1550$ nm).  The mirror pattern consists of a periodic array of holes, whose dimensions are tapered from circles to elliptical shapes with a profile similar to the gap.  Crucially, the band edge of the photonic crystal ``air mode'' associated with the gap unit cell is phase matched with the band edge of the neighboring elliptical hole unit-cell air mode, minimizing radiation loss in the gap region and creating a smooth ``optical potential'' for localized modes  \cite{ref:deotare2009hqf, ref:eichenfield2009mdc, ref:hryciw2013ods}.  The high-$Q_o$ optical mode supported in the gap region has a field distribution shown in Fig.\ \ref{fig:Optics}(a) and is characterized by a mode volume $V_\text{o} \sim 0.3 (\lambda_o/n_{Si})^3$ and radiation loss limited $Q_o \sim 10^4 - 10^6$, depending on the minimum realizable feature size \cite{ref:hryciw2013ods}. The design utilized here is predicted from finite-element simulations ($\text{COMSOL}$) to support a mode with $Q_o \sim 3.5 \times 10^4$.

\begin{table}[t]
\begin{tabular}{c c c c c}
\hline
\hline

Mechanical mode					&			& $T_y$		& $T_z$		&	$C$		\\
\hline
$\omega_m/2\pi$ 			& (MHz)		&	4.9	&	6.4	&	7.7	\\
$m$  							& (fg)		&	427		&	805		&	348 		\\
$Q_m$ (ambient) 			& 			&	21  		&	83 		&	42  		\\
$Q_m$ (vacuum)		& 			&	1800	&	4400	&	2400 	\\
$|z,x|_\text{NF}$ (ambient) & (fm$/\sqrt{\text{Hz}}$) 		&	6.3	&	6.9	&	... \\
$|\tau|_\text{min}$ (ambient) & (N$\,$m$/\sqrt{\text{Hz}}$)		&	$1.2\times 10^{-20}$		&	$1.2\times 10^{-20}$		&	... \\
$|\tau|_\text{min}$ (vacuum) & (N$\,$m$/\sqrt{\text{Hz}}$)		&	$1.3\times 10^{-21}$		&	$1.7\times 10^{-21}$	& ... \\
\hline
\hline
\end{tabular}
\caption{ Split-beam nanocavity mechanical mode properties.}
\label{fig:Parameters}
\end{table}


The split-beam nanocavity supports several cantilever-like mechanical resonances suitable for torque detection, whose displacement profiles,  calculated from simulations and illustrated in Fig.\ \ref{fig:Optics}(b), are characterized by effective mass $m \sim 350--800$ fg and frequency $\omega_m/2\pi \sim 5--8$ MHz (see Table~\ref{fig:Parameters}). The two lowest-frequency modes involve pivoting of the suspended mirror about its support. They are torsional in the $\hat{y}$ and $\hat{z}$ directions and are thus labeled $T_y$ and $T_z$, respectively. The third mode $C$ is an out-of-plane cantilever-like mode of the triply anchored mirror.


The optomechanical properties of the split-beam nanocavities are measured using a dimpled optical fiber-taper waveguide to evanescently couple light into and out of the nanocavity.  The dimple is fabricated by modifying the process in Ref. \cite{ref:michael2007oft} to use a ceramic dimple mold.  Measurements are performed both in ambient conditions and in vacuum.  A tunable laser source is used to measure $T(\lambda)$, with the taper either hovering approximately $300--500$ nm above the nanocavity or touching one of the nanocavity mirrors.  The nanocavity studied here supports an optical mode at $\lambda_o \sim 1530$ nm, with unloaded $Q_o \sim 12\, 000$ due to fabrication imperfections, resulting in a dip in $T(\lambda)$ near $\lambda_o$, as shown in Fig.\ \ref{fig:OptoMech}(a). Optomechanical coupling between this mode and nanocavity mechanical resonances is studied by measuring the rf noise spectrum $\bar{S}_\text{VV}(\lambda,\omega)$ of the optical power transmitted through the fiber taper, using a photoreceiver (New Focus 1811, detector noise 2.5 pW/$\sqrt{\text{Hz}}$) and a real-time spectrum analyzer (Tektronix RSA5106A).  A typical measurement of $\bar{S}_\text{VV}$, with the fiber hovering above the nanocavity and $\Delta\lambda \sim -\delta\lambda_o/2$, is shown on the right of Fig.\ \ref{fig:OptoMech}(a). Three distinct resonances are visible, indicative of the optomechanical transduction of the thermal motion of the $T_y$, $T_z$, and $C$ modes.  The resonances are identified with mechanical modes through comparison of measured and simulated $\omega_m$, and by observing the effect of touching the fiber taper on each of the mirrors. As shown in Fig.\ \ref{fig:OptoMech}(b), when the fiber contacts the anchored (suspended) mirror, the $C$ ($T_y$ and $T_z$) resonance is suppressed,  as it is a resonance of the anchored (suspended) mirror.  


The mechanical displacement sensitivity of these measurements can be calibrated from $\bar{S}_\text{VV}(\omega=\omega_m)$, which is determined by the thermal amplitude of the mechanical resonance \cite{ref:liu2012wcs, ref:krause2012ahm}.   From the measured and calculated mechanical mode properties listed in Table~\ref{fig:Parameters}, the noise-floor displacement resolutions, $|z,x|_\text{NF}$, for the $T_y$ and $T_z$ modes of approximately $6$ and $7$ fm$/\sqrt{\text{Hz}}$, respectively, are measured for $P_i \sim$ 25 $\mu$W. The minimum detectable torque $\tau_\text{min}$ associated with the angular motion $\theta$ of each mechanical mode can be calculated from $\tau = \textbf{r}\times\textbf{F}$ and Eq.\ \eqref{eq:MinSensitivity}. From the mirror length of $7.5~\mu$m and support length of $3~\mu$m, a thermally limited torque sensitivity of the $T_y$ and $T_z$ modes, in ambient conditions, of $\tau_\text{min} \sim 1.2 \times 10^{-20}~\text{N} \, \text{m}/\sqrt{\text{Hz}}$, and on-resonance technical noise floors of $4 - 7 \times 10^{-22}~\text{N} \, \text{m}/\sqrt{\text{Hz}}$, limited by laser noise and photon shot noise, are extracted. This technical noise floor, corresponding to the second term in Eq.\ \eqref{eq:MinSensitivity}, has an effective temperature in the mK range. Measurements are also performed in low vacuum, where the effect of air damping is reduced. The limit imposed by thermal noise, determined by the first term in Eq.\ \eqref{eq:MinSensitivity}, can be reduced by decreasing the mechanical damping of the device. An increase in $Q_m$ of the $T_y$ and $T_z$ modes, from $Q_m^\text{atm} = 21$ and 83 in ambient pressure to $Q_m^{\text{vac}} = 1800$ and 4400 at a relatively low vacuum pressure of 2 Torr, is observed, as shown in Fig.~\ref{fig:OptoMech}(b) and summarized in Table I. For a given set of operating conditions, Eq.~\eqref{eq:MinSensitivity} indicates that this 2-orders-of-magnitude improvement of $Q_m$ will enhance sensitivity by an order of magnitude, resulting in thermally limited $\tau_\text{min} \sim 1.3 \times 10^{-21}$ and $\sim 1.7 \times 10^{-21}~\text{N} \, \text{m}/\sqrt{\text{Hz}}$ at 2 Torr for the $T_y$ and $T_z$ modes, respectively. Note that higher $Q_m$ results in a reduced bandwidth of the mechanical response $\chi(\omega)$ and is not always preferred for practical applications.

The observed torque sensitivity in ambient conditions is higher and wider in bandwidth than previously demonstrated optomechanical torque sensors in vacuum \cite{ref:kim2013nto}.  The observed vacuum sensitivity is an order-of-magnitude improvement compared to previous work \cite{ref:kim2013nto}.  Further improvements in detection sensitivity can be realized through improvements in optical and mechanical properties of the devices. Increasing the fiber-cavity coupling efficiency from the relatively weak coupling demonstrated here ($T_o \sim 0.92 -- 0.98$),  using single-sided coupling via an integrated waveguide \cite{ref:Groblacher2013hec}, for example, would increase $G$ by an order of magnitude. Similarly, increasing $Q_o$ to $4\times10^5$ by more accurately fabricating the split-beam nanocavity designs \cite{ref:hryciw2013ods} will also enhance $G$.  Combining these improvements, torsional sensitivity could reach $10^{-23} -- 10^{-22}  ~\text{N} \, \text{m}/\sqrt{\text{Hz}}$.

\begin{figure}[t]
\epsfig{figure=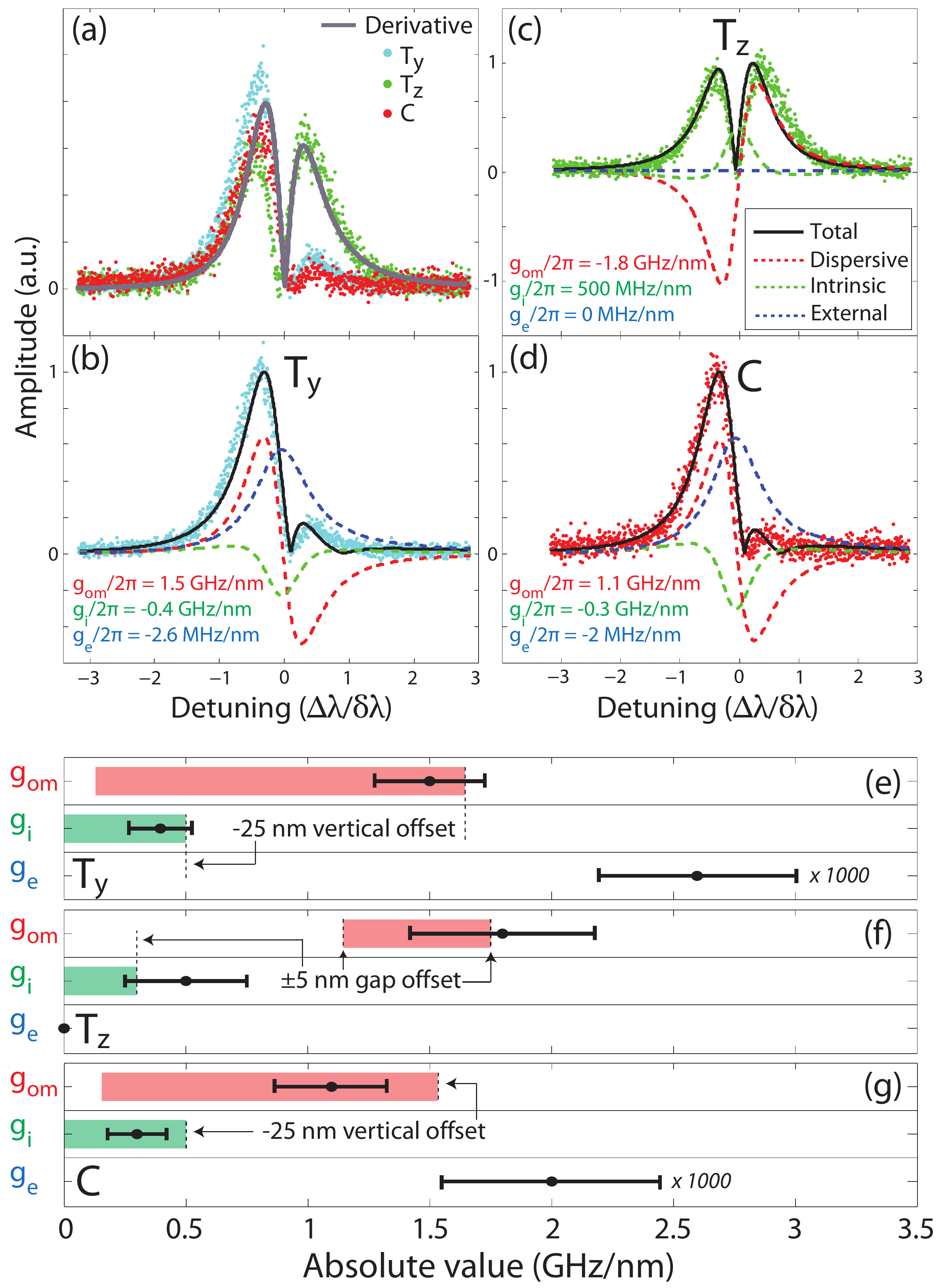, width=1\linewidth}
 \caption{ (a) $\bar{S}_\text{VV}^{1/2}(\lambda,\omega=\omega_m)$ of $C$, $T_{y,z}$ modes. The grey line is scaled $dT(\lambda)/d\lambda$. (b) - (d) Fit (black line) of the optomechanical coupling model to $\bar{S}_\text{VV}^{1/2}(\lambda,\omega=\omega_m)$ of the (b) $T_y$, (c) $T_z$, and (d)  $C$ modes. The dashed colored lines indicate relative contributions from dispersive, external dissipative, and intrinsic dissipative couplings. (e) - (g) Comparison between fit values (black points) and numerically simulated values (shaded regions) of $g_\text{OM}$, $g_i$, and $g_e$. The widths of the colored boxes represent numerically simulated ranges due to fabrication imperfections.}
\label{fig:Dissipative}
\end{figure}


The role of dissipative optomechanical coupling and its effect on torque-detection sensitivity is studied by measuring the wavelength response of the rf spectrum. Examining $\bar{S}_\text{VV}(\lambda, \omega)$ and $T(\lambda)$ in Fig.\ \ref{fig:OptoMech}(a), it is evident that the optomechanical transduction of each of the three mechanical resonances exhibits a unique $\lambda$ dependence. This stems from differing relative contributions of dissipative and dispersive coupling. $S_\text{VV}(\lambda,\omega_m)$ of a purely dispersive cavity-optomechanical system, operating in the unresolved-sideband regime, should follow the slope $|dT/d\lambda|^2$. However, as shown in Fig.\ \ref{fig:Dissipative}(a), $\bar{S}_\text{VV}(\lambda, \omega_m)$ of  $T_y$, $T_z$, and $C$ do not follow $|dT/d\lambda|^2$ and are asymmetric with respect to $\pm \Delta\lambda$. This asymmetry can be characterized by $\zeta^2=\bar{S}_\text{VV}^+/\bar{S}_\text{VV}^-$, where $\bar{S}_\text{VV}^{\pm}$ is the maximum rf sideband signal for $\Delta\lambda \gtrless 0$.  The slight Fano profile of $T(\lambda)$, due to  nanocavity  coupling to higher-order waveguide modes, would result in a $\zeta \sim 0.8$ for purely dispersive optomechanical coupling, and does not explain the observed results. In comparison, $\bar{S}_\text{VV}(\lambda, \omega_m)$ of the out-of-plane $T_y$ and $C$ modes, shown in Figs.\ \ref{fig:Dissipative}(b) and \ \ref{fig:Dissipative} (d), are characterized by $\zeta\sim 0.24$ and $0.19$, respectively. The in-plane mode $T_z$ is characterized by $\zeta \sim 1.1$, as shown in  Fig.\ \ref{fig:Dissipative}(c).

 The relative contribution of each optomechanical coupling process can be estimated by fitting $\bar{S}_\text{VV} (\lambda,\omega_m)$ with a model that includes dispersive, intrinsic dissipative, and external dissipative optomechanical couplings and takes into account the slight Fano shape of $T(\lambda)$. The resulting fits, estimates for $g_\text{OM}, g_i$, and $g_e$, and relative contributions to the optomechanical response, are displayed in Figs.\ \ref{fig:Dissipative}(b) -- (d). The large asymmetry in $T_y$ and $C$ is attributed primarily to external dissipative coupling, resulting from a variation in fiber-nanocavity gap caused by the motion of the mirror, and is quantified by $g_e \sim -2.6$ MHz/nm. The $T_z$ mode is predominantly dispersive, and good agreement with theory is realized with $g_e = 0$.  In order to realize best fits in all of the modes, significant intrinsic dissipative coupling must be included, with $g_i \sim 300--500$ MHz/nm.  Note that in the case of the out-of-plane $C$ and $T_y$ modes, contributions from $g_e$ effectively double the displacement sensitivity of the optomechanical measurement. Furthermore, even for modest $g_e$, the relative contribution to the optomechanical gain $G$ is significant, despite the weak waveguide-nanocavity coupling used here, owing to $G|_\text{max}\propto T_o$ for external dissipative  coupling.

The fit values for $g_\text{OM}$, $g_i$, and $g_e$  were compared with values predicted from numerical simulations, as summarized in Figs.\ \ref{fig:Dissipative}(e)-\ref{fig:Dissipative}(g). A range of values for $g_i$ and $g_\text{OM}$, accounting for uncertainties in device fabrication, is calculated both by directly simulating the optical properties of the nanocavity resonance as a function of mirror displacement and by using perturbation theory (see the Appendix). For $T_z$, the in-plane motion of the suspended mirror contributes to $g_\text{OM}$ and $g_i$. An uncertainty of $\pm$5 nm in the gap size results in the predicted range of $g_\text{OM}$ and $g_i$ shown in Fig.\ \ref{fig:Dissipative}(f). Because of fabrication imperfections unaccounted for in simulations, experimental values can be slightly higher; error bars from the fitting routine, however, fall within the predicted range. For the out-of-plane $T_y$ and $C$ modes, broken vertical symmetry can give rise to significant $g_\text{OM}$ \cite{ref:li2009bat}. Notably, a vertical sagging of the suspended mirror by a plausible offset of 25 nm, as indicated in Figs.\ \ref{fig:Dissipative}(e) and\ \ref{fig:Dissipative}(g), can give rise to $g_\text{OM}$ and $g_i$ values comparable to the fit value for $T_z$.  Note that renormalization of the nanocavity near field by the waveguide can contribute to $g_\text{OM}$, but that this effect is not significant for the operating conditions used here. Finally, the values for $g_e$ extracted from the fits are comparable to the experimentally observed dependence of $\gamma_e$ on waveguide-nanocavity gap \cite{ref:srinivasan2004ofb}. 

Several recent studies have explored the potential for exploiting dissipative optomechanical coupling for applications in the quantum regime \cite{ref:elste2009qni, ref:huang2010rcb, ref:xuereb2011dom, ref:weiss2013qll}.  For many of these proposals, it is desirable to reduce dispersive coupling and maximize dissipative coupling.  This can potentially be achieved in split-beam nanocavities. Simulations indicate (see the Appendix) that internal dissipative coupling of the $T_z$ mode can become dominant if the mirror gap is increased by 50 nm, where $\{g_\text{OM},g_i \} \to \{0,1.5\,\text{GHz/nm}\}$. In the case of the $T_y$ and $C$ modes,  $g_e/g_\text{OM}$  may be increased by reducing the single mirror ``sag" believed to be largely responsible for the appreciable $g_\text{OM}$ measured for these modes. This may be achieved in devices with symmetrically supported mirrors. 

In conclusion, we have observed thermally driven dissipative and dispersive optomechanical couplings in a nanostructure consisting of cantilever nanomechanical resonators integrated directly within an optical nanocavity.  This nanocavity is capable of detecting torque with sensitivity of 1.3$\times 10^{-21}~\text{N} \, \text{m}/\sqrt{\text{Hz}}$ in low vacuum, and 1.2$\times 10^{-20}~\text{N} \, \text{m}/\sqrt{\text{Hz}}$ in ambient conditions. This sensitivity surpasses previously demonstrated optomechanical torque detection in vacuum (4 $\times  10^{-20}~\text{N} \, \text{m}/\sqrt{\text{Hz}}$ in Ref.\ \cite{ref:kim2013nto}) by an order of magnitude, and compares favorably to the performance of magnetic tweezer torque sensors ($\sim10^{-21}~\text{N} \, \text{m}$ in Ref.\ \cite{ref:lipfert2010mtt}). The low-temperature operation of the existing device would allow sensitivity to be improved by up to 2 orders of magnitude for the technical noise floor measured here.  Further optimization of $Q_o$, $Q_m$, $m$, and the optomechanical coupling strength of our device will serve to further reduce technical and thermal noise \cite{ref:forstner2012spc} and to improve the measurement sensitivity and resolution. For example,  $Q_o> 10^5$ \cite{ref:hryciw2013ods, ref:eichenfield2009mdc},  $Q_m > 10^4$ \cite{ref:safavinaeini2013sls} and $g_\text{OM} > 10$ GHz/nm \cite{ref:eichenfield2009mdc} are potentially within experimental reach.  Realizing a device with this combination of performance would provide a path toward further improving torque sensitivity by orders of magnitude, as well as enhancement of the relative strength of dissipative optomechanical coupling. 

This work is supported by the Natural Science and Engineering Research Council of Canada (NSERC), the Canada Foundation for Innovation (CFI) and Alberta Innovates Technology Futures (AITF).  We give special thanks to Hamidreza Kaviani, Brad D. Hauer, and Aashish A. Clerk for helpful discussions. We also thank the staff of the nanofabrication facilities at the University of Alberta and at the National Institute for Nanotechnology for their technical support. \\


\section{APPENDIX}

\subsection{1. Dispersive and dissipative optomechanical coupling} 

Below, we present equations describing the wavelength dependence of the split-beam photonic crystal nanocavity optomechanical response. This model takes into account dissipative and dispersive optomechanical coupling. It also modifies the usual waveguide-cavity temporal coupled mode theory to include indirect coupling between the cavity and the fundamental waveguide mode, mediated by higher-order modes of the waveguide.

The detected optical signal consists of the output field in the fundamental mode of an optical fiber-taper waveguide positioned in the near field of the optical cavity.  The polarization of this mode is chosen to maximize its coupling to the cavity.  The modal output amplitude is 
\begin{equation}\label{eq:to}
t_{o} = s_o + \kappa_{co} a + \kappa_{c+} a,
\end{equation}
where $s_o$ is the input field amplitude and $a$ is the cavity field amplitude.  Coupling from the cavity field into the fundamental fiber-taper mode is described by coupling coefficients $\kappa_{co}$ and $\kappa_{c+}$. 
$\kappa_{co}$  describes coupling from the cavity directly into the fundamental fiber-taper mode, while $\kappa_{c+}$ describes coupling into higher-order modes of the fiber taper that are converted into the fundamental mode along the length of the fiber taper.  Typically, $|\kappa_{c+}| \ll |\kappa_{co}|$, as both the cavity to higher-order mode-coupling process and the fiber-taper higher-order to fundamental mode conversion rates are small.

The cavity-field amplitude is governed by the equation of motion, 
\begin{equation}
\frac{da}{dt} = - \left ( i\Delta + \frac{\gamma_t}{2} \right ) a + \kappa_{oc} s_o
\end{equation}
where  $\Delta = \omega_l - \omega_c$ is the detuning between the input field laser and the cavity frequency, and $\kappa_{oc}$ is the fiber-to-cavity coupling coefficient.   The total cavity-optical loss rate is given by 
\begin{equation}
\gamma_t = \gamma_{i+p} + 2\gamma_e
\end{equation}
where $\gamma_e$ is the coupling rate into the forward (or backward) propagating mode of the fiber taper, and $\gamma_{i+p}$ describes the intrinsic cavity loss and fiber-induced parasitic loss into modes other than the fundamental fiber-taper mode, e.g., scattering into radiation modes and light coupled into higher-order fiber modes  that are not converted into the fundamental waveguide mode within the fiber taper.

In the steady state, $\dot{a} = 0$, and the cavity-field amplitude is:
\begin{equation}
a = \frac{\kappa_{oc}s_{o}}{i\Delta+\frac{\gamma_t}{2}}.
\end{equation}
In the case of a two-port coupler, unitarity requires that $\kappa_{oc} = -\kappa_{co}^* = i \sqrt{\gamma_e}$ for the phase convention chosen in Eq.\ (\ref{eq:to}).  Assuming that the correction due to coupling to higher-order taper modes considered here is small, so that the above relationship still holds, the transmitted field is
\begin{equation}
t_{o} = s_o\left(1 - \frac{\gamma_e+\kappa_{oc}\kappa_{c+}}{i\Delta+\frac{\gamma_t}{2}}\right).
\end{equation}
A key property of the output coupling mediated by the higher-order waveguide mode is that \textit{a priori} the complex phase of $\kappa_{c+}$ is not defined  relative to the phase of $\kappa_{oc}$, as it depends on the modal coupling process between the cavity's coupling region and the fiber taper.  This variable phase leads to a non-Lorentzian cavity response, as seen by writing $\kappa_{c+} = \kappa_{+}^r + i \kappa_{+}^i$, where $\kappa_{+}^r$ and $\kappa_{+}^i$ are both real, and calculating the normalized taper transmission
\begin{equation}
T =  
\frac{\Delta^2 + \left(\frac{\gamma_{i+p}}{2}\right)^2  + 2\sqrt{\gamma_e}\kappa_{+}^r\Delta - \sqrt{\gamma_e}\kappa_{+}^i\gamma_{i+p}}{\Delta^2 + \left(\frac{\gamma_t}{2}\right)^2},
\label{Transmission}
\end{equation}
where we have only kept terms to lowest order in $\kappa_{+}^{i,r}$. For weak fiber-cavity coupling, $\gamma_e \ll \gamma_{i+p}$, the last term in the denominator can be ignored, and
\begin{equation}
T \sim  
\frac{\Delta^2 + \left(\frac{\gamma_{i+p}}{2}\right)^2 + C_f\gamma_e\Delta}{\Delta^2 + \left(\frac{\gamma_t}{2}\right)^2},
\label{eq:T_fano}
\end{equation}
where $C_f = 2\kappa_+^r/\sqrt{\gamma_e}$ represents a Fano modification to the cavity response mediated by the higher-order fiber-taper modes and is expected to be small.
 
The optomechanical response of our cavity can be modeled by considering the dependence of the parameters in Eq.\ (\ref{eq:T_fano})  on the mechanical state $x$ of the mechanical resonance of interest.  In the unresolved-sideband regime, where the mechanical frequency is small compared to the optical linewidth $\omega_m \ll \gamma_\text{t}$, the fiber transmission adiabatically follows the mechanical oscillations.  The amplitude of the optical oscillations for a given mechanical displacement amplitude $dx$ is 
\begin{equation}
\frac{dT}{dx}(\Delta) = \left | g_\text{OM} \frac{\partial T}{\partial \Delta} + g_i \frac{\partial T}{\partial \gamma_i} + g_e \frac{\partial T}{\partial \gamma_e} \right |
\label{eq:dTdx}
\end{equation}
where $g_\text{OM} = d\omega_o/dx$ is the dispersive optomechanical coupling coefficient~\cite{ref:eichenfield2009mdc}, $g_i = d\gamma_i/dx$ is the intrinsic dissipative coupling coefficient, and $g_e =  d\gamma_e/dx$ is the external dissipative coupling coefficient. The derivatives of Eq.\ (\ref{eq:T_fano}) are 
\begin{align}
 \frac{\partial T}{\partial \Delta} &= \frac{2\Delta(1-T) + \gamma_e C_f}{\Delta^2 + \left(\gamma_t/2\right)^2}\\
	\frac{\partial T}{\partial \gamma_i} &= \frac{\gamma_\text{i+p} - T (\gamma_{i+p} + 2\gamma_e)}{\Delta^2 + \left(\gamma_t/2\right)^2}\\
 \frac{\partial T}{\partial \gamma_e} &= \frac{ - 2\gamma_t T + \Delta C_f}{\Delta^2 + \left(\gamma_t/2\right)^2}.
\label{eq:dTdgi}
\end{align}

\begin{figure}[h]
\begin{center}
\epsfig{figure=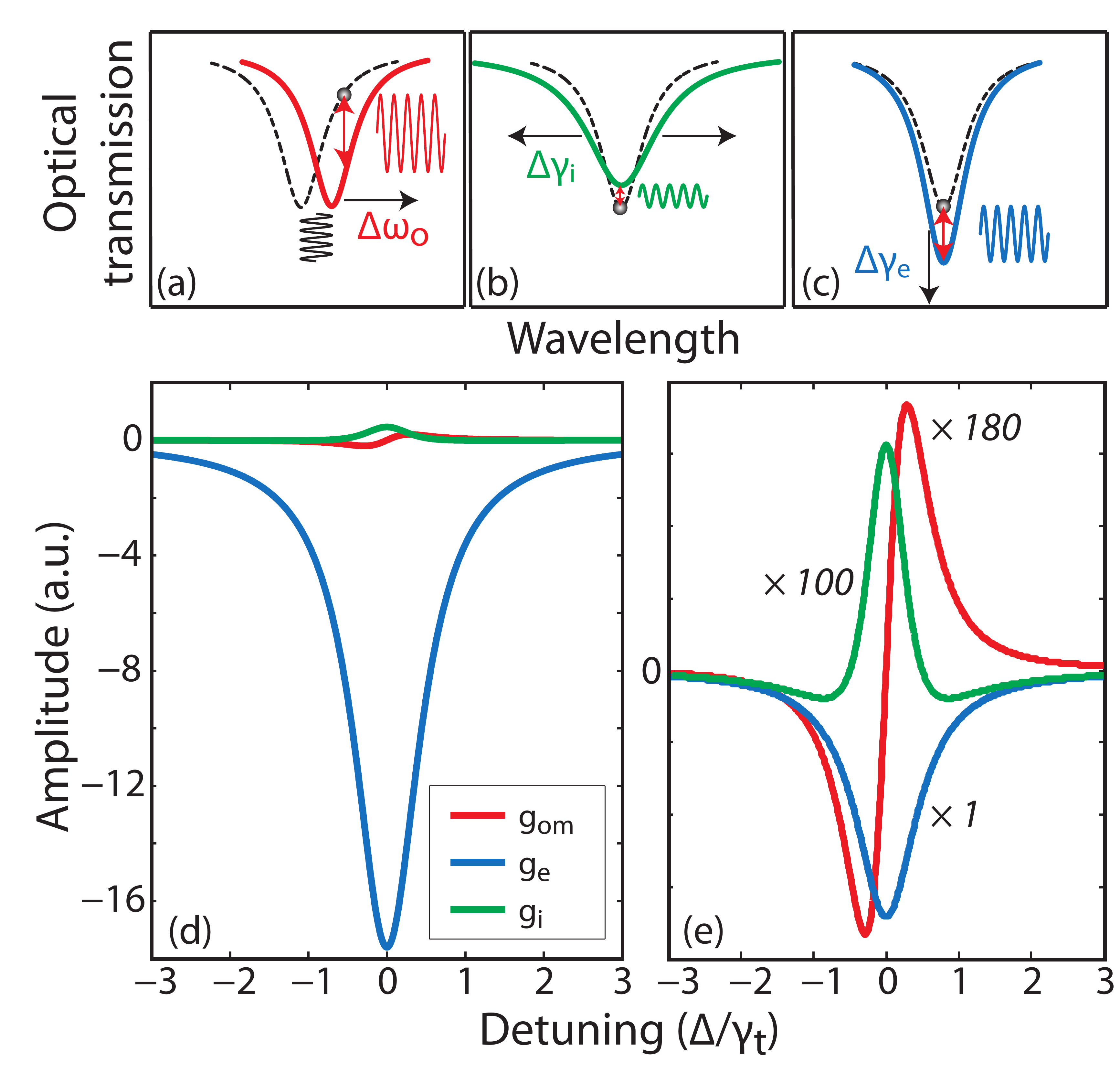, width=1\linewidth}
 \caption{Change in optical transmission due to (a) dispersive, (b) intrinsic dissipative, and (c) external dissipative optomechanical couplings, respectively. (d) Relative strength of the three contributions: $\frac{\partial T}{\partial \Delta}$, $\frac{\partial T}{\partial \gamma_i}$, and $\frac{\partial T}{\partial \gamma_e}$. (e) Comparative strength when contributions are brought to similar amplitudes. The fano modification is omitted for display purposes.}
\label{fig:Contributions}
\end{center}
\end{figure}

The influence of the derivatives of the optical resonance is depicted in Figs.~\ref{fig:Contributions}(a) - \ref{fig:Contributions}(c).  The derivatives with $C_f = 0$ are plotted in Fig.~\ref{fig:Contributions}(d) and scaled in Fig.~\ref{fig:Contributions}(e). A few observations can be made. Our device is undercoupled; $\gamma_e \ll \gamma_i$. Thus, the external dissipative coupling has a larger influence on the line shape due to the fact that the decay rate into the fiber is much smaller ($\gamma_e \approx $ 1 GHz) compared to the cavity linewidth ($\gamma_i \approx$ 30 GHz). We can quantitatively compare this by evaluating the peak amplitude of the derivatives~\cite{ref:krause2012ahm}: 
\begin{align}
\frac{\partial T}{\partial \Delta} \bigg|_\text{max} &= \frac{dT}{d\Delta}(\Delta = \frac{\gamma_t}{2}) = (1-T_{\text{o}}) \frac{Q_o}{\omega_{o}}\\
\frac{\partial T}{\partial \gamma_i} \bigg|_\text{max} &= \frac{dT}{d\gamma_i}(\Delta = 0) = 4 (1-T_o) \frac{Q_o}{\omega_{o}}\\
\frac{\partial T}{\partial \gamma_e} \bigg|_\text{max} &= \frac{dT}{d\gamma_e}(\Delta = 0) = -8 T_o \frac{Q_o}{\omega_{o}}
\end{align}
where $T_o = \gamma_{i+p}^2/\gamma_{t}^2$ is the transmission at optical resonance $\omega_o$ and $Q_o = \omega_o/\gamma_{t}$ is the optical quality factor. Because of small fiber-cavity coupling ($T_o \approx 1$), a change in the fiber coupling has a larger influence on the transmission near resonance such that $\frac{\partial T}{\partial \gamma_e}$ dominates over the other terms, as seen in Fig.~\ref{fig:Contributions}(d). Hence, a small value of $g_e$ has a greater effect on the change in transmission than large values of $g_i$ and $g_\text{OM}$. With a single peak at zero detuning for $\frac{\partial T}{\partial \gamma_i}$ and $\frac{\partial T}{\partial \gamma_e}$, it is possible to exploit optomechanics at resonance, provided dissipative coupling is stronger than dispersive. By careful mixing of the coupling rates, optomechanical cooling is also possible, at least in theory~\cite{ref:weiss2013qll}. Cooling arises through quantum noise interference between dispersive and dissipative couplings, with the best results occurring when operating with external dissipative coupling~\cite{ref:krause2012ahm}.

\subsection{2. Power spectral density and thermomechanical calibration} 

The transduction of the resonator mechanical motion to a photodetected electronic signal, the subsequent analysis of the electronic power spectral density, and the relationship between this power spectral density and the optomechanical coupling coefficients of the device are given below.  In the setup used here, a real-time electronic spectrum analyzer (RSA) samples the time-varying voltage, $V(t) = V_\text{OM}(t) + V_n(t)$, generated by a photoreceiver input with the optical field transmitted through the fiber taper.  For a given input power $P_i$ and operating wavelength $\lambda$, the optomechanical contribution $V_\text{OM}(t)$ to this signal is given by
\begin{equation}
V_\text{OM}(t) = \eta g_{ti} P_i T(\lambda, x(t))
\end{equation}
where  $g_{ti}$ is the photoreceiver transimpedance gain (40 000 V/W assuming a 50-$\Omega$ load), and $\eta$ accounts for loss between the detector and fiber-taper output.  Technical fluctuations $V_{n}(t)$ arise from optical, detector, and electronic measurement noise.

In general, the fiber-taper transmission $T$ varies, depending on the general displacement $x$ of the nanocavity mechanical resonator and the effect of $x$ on the optical response of the fiber-coupled nanocavity. Here, $x(t)$ describes the thermally driven fluctuations of the nanocavity mechanical resonator.  The device considered in this paper is operating in the sideband-unresolved regime ($\omega_m \ll \gamma_t$), where the nanocavity field can ``follow'' the mechanical oscillations, allowing us to write
\begin{equation}\label{eq:sig}
V_\text{OM}(t) = \eta g_{ti} P_i \left(T_o + \frac{dT(\lambda)}{dx} x(t)\right).
\end{equation}
The RSA demodulates $V(t)$ and outputs $IQ$ time-series data, $V_{IQ}(t) = I(t) - iQ(t)$, where $I(t) = \cos(\omega_c t) V(t)*h(t)$ and  $Q(t) = \sin(\omega_c t) V(t)*h(t)$.  Here, $\omega_c$ is the demodulation frequency, and $h(t)$ is a low-pass antialiasing filter, whose span is determined by the sampling rate (up to 40 MHz). The Fourier transform of  the $IQ$ data is related to the input spectrum by $\bar{V}_{IQ}(\omega) =  V(\omega + \omega_c)H(\omega)$.  Note that a scaling factor is built into $H(\omega)$ to ensure that  $|\bar{V}_{IQ}(\omega - \omega_c)|^2$ can be accurately treated as a single-sided (positive frequency) representation of the symmetrized input power spectrum.

The two-sided  power spectral density of the optomechanical contribution to the input signal is given by 
\begin{equation}
S_\text{VV}^\text{OM}(\omega) = |V_\text{OM}(\omega)|^2/\Delta t,
\end{equation}
where $\Delta t$ is the acquisition time of the RSA time series, and $V(\omega) = \int_0^{\Delta t} dt\, e^{-i\omega t} V(t)$. For clarity, the dc component is ignored in the following analysis.  Using Eq.\ (\ref{eq:sig}), $S_\text{VV}^\text{OM}$ can be related to the stochastically varying displacement $x(t)$ of the nanocavity
\begin{align}
S_\text{VV}^\text{OM}(\omega) 	&= \left(\eta g_{ti} P_i \frac{dT(\lambda)}{dx}\right)^2\frac{1}{\Delta t}\left|\int_0^{\Delta t}dt \, e^{-i\omega t}  x(t) \right|^2\\
 			&= \frac{G^2}{\Delta t} \int_0^{\Delta t} dt' \int_{-t'}^{\Delta t-t'} dt \, e^{-i\omega t} x^*(t'+t) x(t'),
\end{align}
where $G = \eta g_{ti} P_i\, dT/dx$ describes the detector and the optomechanical response.
The stationary nature of $x(t)$, i.e., $\left\langle x^*(t+t')x(t)\right\rangle = \left\langle x^*(t')x(0)\right\rangle$ for measurement time $\Delta t \gg 2\pi/\gamma_{m}$, allows us to write the above equation as
\begin{align} 			
S_\text{VV}^\text{OM}(\omega) 	&= G^2 \int_{0}^{\Delta t} dt' \, e^{-i\omega t'} \langle x^*(t+t') x(t) \rangle\\
  					 &= G^2(\lambda) S_{xx}(\omega),
\end{align}
where $S_{xx}(\omega)$ is the displacement noise spectral density of the mechanical resonator.  The total single-sided power spectral density measured by the RSA is
\begin{equation}
\bar{S}_\text{VV}(\lambda,\omega) = G^2(\lambda) \bar{S}_{xx}(\omega) + \bar{S}_\text{VV}^n(\lambda, \omega),
\label{FreqSpp}
\end{equation}
where the contribution from the technical noise is labeled $\bar{S}_\text{VV}^n(\lambda,\omega)$. Note that the spectral density can also be expressed as true power over a load resistance $Z$ such that $\bar{S}_p = \bar{S}_\text{VV}/Z$ in units of W/Hz or dBm/Hz. 

The displacement noise of a thermally excited mechanical mode $m$ can be derived from the fluctuation-dissipation theorem \cite{ref:cleland2002npn} and is given by
\begin{equation} 
\bar{S}_{xx}(\omega) = \frac{4k_BT_e\omega_m}{Q_m} \frac{1}{m [(\omega^2-\omega_m^2)^2 + (\frac{\omega \omega_m}{Q_m})^2]},
\end{equation}
where $k_B$ is Boltzmann's constant, $T_e$ = 300 K for the experiments conducted here, and $m$ is the effective mass as defined in Ref.~\cite{ref:eichenfield2009mdc}. On mechanical resonance, $\omega = \omega_m$, the power spectral density becomes
\begin{equation}\label{eq:S_vv_full}
\bar{S}_\text{VV} (\lambda) \big |_{\omega = \omega_m} = G^2(\lambda)\frac{4 x_{\text{rms}}^2 Q_m}{ \omega_m} + \bar{S}_\text{VV}^n(\lambda),
\end{equation}
where $x_{\text{rms}} = \langle x^2 \rangle ^{\frac{1}{2}} = \sqrt{k_BT_e/m\omega_m^2}$ is the mean thermal displacement.   In the work presented here, backaction effects are not significant, and the mechanical parameters of the device $Q_m$ and $\omega_m$ are independent of $\lambda$ and $P_i$. These parameters can be extracted by fitting Eq.\ \eqref{FreqSpp} to the measured spectrum for a given $\lambda$ (usually chosen to maximize $\bar{S}_\text{VV}/\bar{S}_\text{VV}^n$). 

The dispersive and dissipative optomechanical coupling coefficients, $g_\text{OM}$, $g_i$, and $g_e$ can be extracted from the experimental data by fitting the $\lambda$ dependence of  $S_\text{VV}(\lambda,\omega = \omega_m)$ to Eq.\ (\ref{eq:S_vv_full}).  The interplay between optomechanical coupling mechanisms, as well as the line shape of the nanocavity optical resonance, is captured by $G(\lambda)$ via its dependence on $dT/dx$, as described theoretically in Eqs.\ (\ref{eq:dTdx})--(\ref{eq:dTdgi}).

The noise-floor displacement resolution $x_\text{NF}$ of the optomechanical transduction for a given set of operating conditions can be determined from Eq.~\eqref{eq:S_vv_full}, from which the displacement resolution can be calibrated. This method, widely employed by other researchers (see Ref. \cite{ref:hauer2013gpt} and references therein), is used to calibrate the $y$ axis in Fig.\ \ref{fig:OptoMech}(b). The torque sensitivity is calculated based on the on-resonance spectral signal and Eq.~\eqref{eq:MinSensitivity} in the main text.  It can also be theoretically calculated using the effective moment of inertia $I_\text{eff} = r^2 m_\text{eff}$, where $r$ is the distance from the axis of rotation to the position of maximum displacement, and the thermally limited torque sensitivity $\tau_\text{th} = \sqrt{4k_B T_e w_\text{m} I_\text{eff}/Q_m}$. Both methods arrive at the same result.

\subsection{3. Numerical simulations of optomechanical coupling} 

\begin{figure*}[!ht]
\epsfig{figure=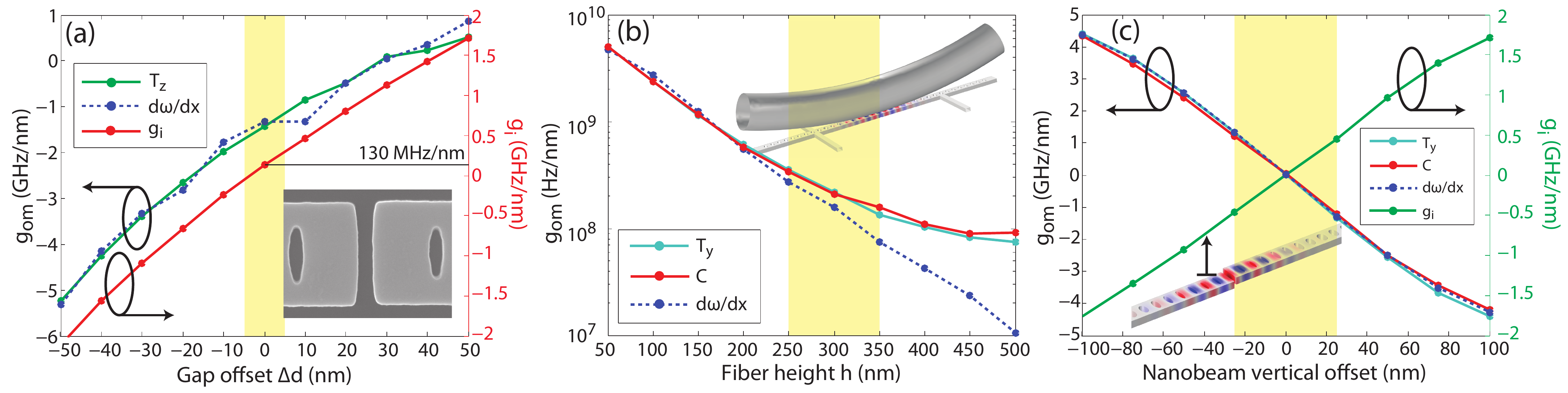, width=1\linewidth}
 \caption{Numerical simulation of the coefficients $g_\text{OM}$ and $g_i$ from finite-element simulations ($COMSOL$). The dashed line $g_\text{OM}$ data are calculated directly from $\omega(x)$. $g_i$ calculated from $\gamma_i(x)$.  All other data points are calculated perturbatively. (a) Dependence of $g_\text{OM}$ (left axis) and $g_i$ (right axis) of the torsional mode $T_z$ on the variation in the gap size $d$. The shaded area corresponds to uncertainty in the gap ($\pm$ 5 nm) due to fabrication tolerances and scanning electron micrograph image resolution. (b) Fiber-induced dispersive optomechanical coupling coefficient $g_\text{OM}$ of the out-of-plane modes $T_y$ and C. The shaded area indicates the uncertainty of the fiber height $h$. Larger $g_\text{OM}$ values at higher $h$ are due to limited finite-element resolution. (c) Dependence of $g_\text{OM}$ (left axis) and $g_i$ (right axis) for out-of-plane modes on the vertical offset  of the suspended mirror.  The shaded area corresponds to a region of uncertainty ($\pm$ 25 nm) in the vertical position of the suspended mirror due to postfabrication stresses and substrate effects.  }
\label{fig:RateSimulations}
\end{figure*}

Numerical simulations are performed to predict the dispersive $g_\text{OM}$ and dissipative internal $g_i$  optomechanical coupling coefficients for each of the mechanical modes of the split-beam nanocavity. In addition to predicting $g_\text{OM}$ of the torsionally actuated $T_z$ mechanical mode of the split-beam nanocavity, these simulations assess the effect on the optomechanical coupling of fabrication imperfections and the presence of the optical fiber taper in the nanocavity near field. All simulations are performed using $COMSOL$ finite-element software to calculate the mechanical and optical mode field distributions and properties $\omega_o$, $\omega_m$, $m$, and $\gamma_i$.  Dispersive optomechanical coupling coefficients $g_\text{OM}$ are calculated using perturbation theory, as by Eichenfield et al.\ \cite{ref:eichenfield2009mdc}, and directly from $g_\text{OM} = d\omega_o/dx$ where $\omega_o(x)$ is the optical mode frequency as a function of mechanical displacement $x$.  Dissipative $g_i$ are calculated directly from $d\gamma_i/dx$.  Simulations are performed for a range of device dimensions consistent with our observed fabrication tolerances.

The in-plane motion of the $T_z$ mode modulates the split-beam gap width $d$, resulting in a large dispersive optomechanical coupling. Figure \ref{fig:RateSimulations}(a) shows $g_\text{OM}$ for this mode as a function of an offset $\Delta d$ away from the nominal value of $d=60$ nm. For $\Delta d=0$, $g_\text{OM}=-1.5$ GHz/nm is predicted, using both perturbation and direct $d\omega/dx$ calculation techniques. If $d$ is not optimized, small displacements of $T_z$ will also modify $\gamma_i$. For our best estimate of the gap size, simulations predict $g_i$=130 MHz/nm; however within the uncertainty in position this value can vary.

Because of the different vertical symmetry of the nanocavity optical mode, and the displacement fields of out-of-plane modes $T_y$ and C,   their optomechanical coupling coefficients $g_\text{OM}$ and $g_i$ are expected to be 0.  However, the vertical symmetry of the optical mode is broken in two ways in the device studied here.  Interactions between the nanocavity evanescent field and the optical fiber taper, for small fiber-taper height $h$ above the nanocavity surface, modify the effective refractive index of the nanocavity.  This effect is described by an $h$-dependent $g_\text{OM}(h)$ that can reach the GHz/nm range, as shown in Fig.\ \ref{fig:RateSimulations}(b).  Fabrication imperfections in the device and the fabrication process can also break vertical symmetry. Notably, offset bending between the two mirrors can arise due to differential internal stresses in each beam and stiction forces due to the proximity of the substrate. This is referred to as ``sagging'' in the discussion below. Figure \ref{fig:RateSimulations}(c) illustrates the effect of sagging in the suspended mirror, resulting in an offset in the $z$ direction with respect to the anchored mirror.   Broken vertical symmetry also manifests in nonzero intrinsic dissipative optomechanical coupling for the out-of-plane modes. Sagging of the suspended mirror shifts the nanocavity cavity mode away from the minimum intrinsic loss $\gamma_i$, resulting in $g_i = d\gamma_i/dx < 0$, as shown in Fig.~\ref{fig:RateSimulations}(c).

Numerical simulations for the fiber-cavity external coupling coefficient $g_e$ were inconclusive. For typical values of $g_e\sim2$ MHz/nm \cite{ref:srinivasan2004ofb, ref:li2009rco, ref:anetsberger2011coc}, the change in $Q_o$ for our cavity ($Q_o\sim12 000$) due to a change $h$ of 100 nm would be in the order of $\Delta Q\sim100$, which is below the uncertainty of our numerical simulations for this specific device.

\bibliography{nano_bib}

\end{document}